\documentclass[12pt,a4paper]{article}
\usepackage{type1cm,amsmath,hangcaption,graphicx,indentfirst}
\usepackage[psamsfonts]{amssymb}
\numberwithin{equation}{section}

\begin{document}
\begin{titlepage}

 \renewcommand{\thefootnote}{\fnsymbol{footnote}}
\begin{flushright}
 \begin{tabular}{l}
 DESY 07-085\\
 NSF-KITP-07-140\\
 \end{tabular}
\end{flushright}

 \vfill
 \begin{center}
 \font\titlerm=cmr10 scaled\magstep4
 \font\titlei=cmmi10 scaled\magstep4
 \font\titleis=cmmi7 scaled\magstep4
 \centerline{\titlerm Closed String Tachyons on AdS Orbifolds}
 \vskip .3 truecm
\centerline{\titlerm  and Dual Yang-Mills Instantons}

 \vskip 2.5 truecm

\noindent{ \large Yasuaki Hikida$^a$\footnote{E-mail:
yasuaki.hikida@desy.de}
and Norihiro Iizuka$^b$\footnote{E-mail: iizuka@kitp.ucsb.edu}}
\bigskip

 \vskip .6 truecm
\centerline{\it $^a$DESY Theory Group, Notkestrasse 85, D-22603 Hamburg, Germany}
\bigskip
\centerline{\it $^b$Kavli Institute for Theoretical Physics, }
\centerline{\it University of California, Santa Barbara, CA 93106-4030, USA}

 \vskip .4 truecm

 \end{center}

 \vfill
\vskip 0.5 truecm

\begin{abstract}

We study the condensation of localized closed string tachyons
on AdS orbifolds both from the bulk and boundary theory viewpoints.
We first extend the known results for $AdS_5/{\mathbb Z}_k$ to
$AdS_3/{\mathbb Z}_k$ case, and
we proposed that the $AdS_3/{\mathbb Z}_k$ decays into
$AdS_3/{\mathbb Z}_{k'}$ with $k' < k$. From the bulk viewpoint, we obtain a time-dependent
gravity solution describing the decay of AdS orbifold numerically.
From the dual gauge theory viewpoint, we calculated the
Casimir energies of gauge theory vacua and it is found that their values are
exactly the same as the masses of dual geometries, even though they are 
in different parameter regimes of 't Hooft coupling.
We also consider $AdS_5$ orbifold.
The decay of $AdS_5/{\mathbb Z}_k$ is dual to the transition
between the dual gauge theory vacua on $R_t \times S^3/{\mathbb Z}_k$, 
parametrized by different holonomies along the orbifolded spatial cycle.  
We constructed the instanton solutions describing the transitions
by making use of instanton solutions on $R_t \times S^2$.

\end{abstract}
\vfill
\vskip 0.5 truecm

\setcounter{footnote}{0}
\renewcommand{\thefootnote}{\arabic{footnote}}
\end{titlepage}

\newpage

\section{Introduction}
\label{Intoduction}

Closed string tachyons are very important spectra in string theory since they  
signal the decay of background space-time geometry into others, thus
it is likely that these tachyon condensations play crucial roles in
quantum gravity. 
However, compared with the developments of open string tachyon condensation \cite{Sen:2004nf},  
the dynamics of closed string tachyon condensation is not so well-understood, despite with the  
pioneer work done for the localized closed string tachyon in 
ALE spaces \cite{APS}.\footnote{
For a review, see, e.g., \cite{Headrick:2004hz}.}
On the other hand, significant progresses have been made on AdS/CFT 
correspondence \cite{Maldacena}, which gives a useful tool to understand 
the nature of quantum gravity from the well-defined gauge theory
point of view.
Therefore, we could expect to obtain a lot of insights by
investigating the fate of localized closed string tachyons 
if we embed them into asymptotically AdS spaces.

First let us recall the known results on the condensation of
localized closed string tachyon.
For instance, superstrings on ${\mathbb C}/{\mathbb Z}_k$
were analyzed in \cite{APS}, where the orbifold is constructed
as the two dimensional plane divided by the symmetry of
$2\pi/k$ rotation. The target space is a cone and closed
strings can be localized at the tip of the cone.
Suppose we choose odd $k$  and anti-periodic boundary conditions for fermions,
then the spectrum of closed strings includes no bulk tachyons but
localized tachyons.
An amazing conjecture was given in \cite{APS} that
a localized tachyon condensation leads ${\mathbb C}/{\mathbb Z}_k$ 
into ${\mathbb C}/{\mathbb Z}_{k'}$ with odd $k' < k$
and finally the system ends up with the stable supersymmetric flat vacuum.
The conjecture was confirmed by various ways, such as,
D-brane probes, worldsheet RG-flow, and so on.
In particular, exact gravity solutions describing
the decay of ${\mathbb C}/{\mathbb Z}_k$ were obtained
in \cite{GH,Headrick}.

In this paper we investigate the condensation of localized 
closed string tachyons in superstring theory on 
$AdS_{d+1}/{\mathbb Z}_k$ with $d=2,4$,
which is constructed by the $d+1$ dimensional AdS space with
the identification of $2\pi/k$ rotation.
The orbifold procedure gives rise to a fixed point at the center, 
and we can construct systems with no bulk tachyons but with
tachyons localized at the fixed point by choosing proper spin 
structures as for ${\mathbb C}/{\mathbb Z}_k$ case.
Considering localized tachyon condensations in asymptotic AdS
spaces, we can make intriguing observations which cannot
be seen for the cases of flat space orbifolds.
Due to the warped factor, the effect of closed string 
tachyons is localized around the fix point of AdS orbifold,
and hence the back-reaction by tachyons induces only normalizable modes but not
non-normalizable modes. This implies that the localized tachyon 
condensation changes only bulk physics but not the boundary conditions.
This should be compared to the 
${\mathbb C}/{\mathbb Z}_k$ case, where the boundary conditions 
are altered through the tachyon condensation. 
In fact, it was argued in \cite{Hikida} that the localized tachyon 
condensation of $AdS_5/{\mathbb Z}_k$ deforms the geometry into 
so-called Eguchi-Hanson soliton \cite{CM1,CM2} with the same 
boundary conditions as for the AdS orbifold. 
This fact is actually very important since in an asymptotically 
AdS space we can deal with all the geometries with the same boundary 
condition at the same time.
For instance, we can discuss the thermal phase structure of 
gravity theory in $AdS_5/{\mathbb Z}_k$ \cite{Hikida}.

Utilizing the AdS/CFT correspondence, we can discuss the 
condensation in terms of dual gauge theory. 
In the global coordinates, the boundary of 
$AdS_{d+1}$ is given by $R_t \times S^{d-1}$, where $R_t$ and
$S^{d-1}$ denote the time direction and the $d-1$ dimensional sphere, respectively.
Since the orbifold action acts also on the boundary of $AdS_{d+1}$,
the dual gauge theory is defined by the orbifold of gauge theory on 
$R_t \times S^{d-1}$. One of the important facts for the orbifold gauge
theory is that the theory has many vacua labeled by the holonomy
matrix along its non-trivial cycle. 
It is natural to propose that the condensation of localized tachyon
is dual to the assignment of non-trivial holonomy, because it is known 
that the deformation by normalizable modes corresponds to giving 
expectation values to dual operators. 
Notice that this is related to the fact that the localized tachyon 
condensation does not change the boundary condition as mentioned before.
In this way, we can analyze the tachyon condensation
in terms of dual gauge theory as a transition between different vacua.
In particular, the Casimir energies for the vacua of dual gauge theory
on $R_t \times S^3$ were computed in \cite{Hikida}, and it was found
that they reproduce quite well the masses of dual geometries.

One of the purpose of this paper is to extend the analysis on the 
localized tachyon condensation of $AdS_5/{\mathbb Z}_k$ \cite{Hikida} 
into the case of $AdS_3/{\mathbb Z}_k$.
Up to now only the comparison between the static geometries deformed
by localized tachyon condensation and the vacua of dual gauge theory 
has been done. So we would like to investigate the dynamics
of localized tachyon condensation both from the bulk and
boundary points of view. 
In section \ref{ads3} we study the localized tachyon
condensation on $AdS_3/{\mathbb Z}_k$ with odd $k$.
First we observe that the geometry after the tachyon
condensation is $AdS_3/{\mathbb Z}_{k'}$ with odd $k' <k$
and the final geometry is given by $AdS_3$ without orbifolding.
Next we study the decay of $AdS_3/{\mathbb Z}_k$ following
a dilaton pulse, which is induced by a localized tachyon condensation.
We solve numerically the Einstein-dilaton equations 
in order to obtain a time-dependent gravity solution describing the decay.
The gauge theory dual to $AdS_3/{\mathbb Z}_k$ is defined
on $R_t \times S^1/{\mathbb Z}_k$, and the holonomy matrix along 
the spatial cycle leads to different vacua.
The Casimir energies for the vacua of the gauge theory are 
computed and exact matches are found between the Casimir energies 
and the masses of dual geometries.
In section \ref{ads5} we first review the result of \cite{Hikida},
where it was discussed that $AdS_5/{\mathbb Z}_k$ decays into 
Eguchi-Hanson solution \cite{CM1,CM2}. The dual gauge theory is 
defined on $R_t \times S^3/{\mathbb Z}_k$, and the vacua with
different holonomy correspond to the different geometries
deformed by localized tachyon condensation.
In subsection \ref{gauge5} we construct instanton solutions of 
the orbifold gauge theory on $R_t \times S^3/{\mathbb Z}_k$
with the help of instanton solutions on $R_t \times S^2$ \cite{PhD,Lin}.
The instanton solutions interpolate different vacua,
which are dual to the transitions between different geometries.

\section{Decay of $AdS_3/{\mathbb Z}_k$}
\label{ads3}

We start from extending the results of \cite{Hikida} into 
$AdS_3/{\mathbb Z}_k$ case. We find that the decay 
process of $AdS_3/{\mathbb Z}_k$ is quite similar to the one of 
${\mathbb C}/{\mathbb Z}_k$, namely, $AdS_3/{\mathbb Z}_k$ 
decays into $AdS_3/{\mathbb Z}_{k'}$ with $k' < k$ and ends 
up with pure $AdS_3$. 
Making use of this similarity, we analyze in subsection
\ref{gravity} the dynamics of localized tachyon 
condensation.
A localized tachyon condensation leads to a dilaton pulse,
which travels from the fixed point into the AdS boundary.
The back-reaction of this dilaton pulse induces the decay
of  $AdS_3/{\mathbb Z}_k$ into $AdS_3/{\mathbb Z}_{k'}$ with $k' < k$.
We try to find a time-dependent gravity solution describing the decay in a
numerical way.
The boundary of $AdS_3/{\mathbb Z}_k$ is given by 
$R_t \times S^1/{\mathbb Z}_k$, and the dual gauge theory
is defined on the boundary.
In subsection \ref{gauge3} we define the dual gauge theory
and find the spectrum for various vacua with non-trivial holonomy. 
The Casimir energies of the vacua are computed, and they are shown to 
match precisely with the masses of dual geometries.


\subsection{The deformed geometries after the tachyon condensation}
\label{final3}

Let us consider type IIB superstring theory on 
$AdS_3 \times S^3\times T^4$.
In the global coordinates, the metric of $AdS_3$ is given by
\begin{align}
 ds^2 &= \frac{dr^2}{g(r)} 
   - g(r) dt^2 + r^2 d \theta^2 ~,
 &g(r) &= 1 + \frac{r^2}{l^2} ~.
 \label{AdS3}
\end{align}
The orbifold of $AdS_3$ can be constructed from the
identification of $\theta \sim \theta + 2 \pi /k$,
which gives rise to a fixed point at $r=0$.
Following the arguments of \cite{APS} on ${\mathbb C}/{\mathbb Z}_k$, 
we can construct the configuration with no bulk tachyons and only
tachyons localized at $r=0$ by assuming an odd integer $k$ and 
anti-periodic boundary conditions for fermions. 
Here we have used the fact that local properties do not depend 
on the curvature of AdS space.
{}From the experience of the flat orbifold case,
it is natural to guess that the condensation of localized tachyon 
deforms the orbifold $AdS_3/{\mathbb Z}_k$ into 
$AdS_3/{\mathbb Z}_{k'}$ with odd $k' < k$ and finally into 
the stable supersymmetric vacuum with $AdS_3$.

It is very difficult to prove this conjecture since we do not fully 
understand the localized closed string tachyon.
However, it is possible to obtain several supports for this 
conjecture if we utilize the properties of asymptotic AdS space.
Suppose that the potential of localized tachyonic modes has 
various minima at finite configurations. Then the tachyon 
condensation leads to the deformation of normalizable modes,
which ends up with a deformed geometry with the same boundary condition.
In fact, we can show that $AdS_3/{\mathbb Z}_{k}$ can be deformed 
into $AdS_3/{\mathbb Z}_{k'}$ with odd $k' < k$ without changing 
the boundary behavior. 
Moreover, we observe that the background mass decreases 
as $k'$ becomes small and the smallest mass is given by
$AdS_3$ within the configurations with fixed boundary condition.

In order to describe the $AdS_3/{\mathbb Z}_{k'}$ geometry
with a fixed boundary condition, 
it is not appropriate to use the metric \eqref{AdS3} 
with the identification $\theta \sim \theta + 2\pi/{k'}$
since the boundary condition manifestly depends on the choice of $k'$. 
Instead we use the following metric as 
\begin{align}
 ds^2 &= \frac{dr^2}{g(r) f(r)} 
   - g(r) dt^2 + r^2 f(r) d \theta^2 ~,
 &g(r) &= 1 + \frac{r^2}{l^2} ~, &f(r) &= 1 - \frac{a^2}{r^2} ~.
 \label{AOS}
\end{align}
The period of $\theta$ is set as $\theta \sim \theta + 2 \pi / k$
and the parameter $a$ is related to $k' (<k)$ as
\begin{align}
a^2 &= l^2 \left( K^2  - 1 \right) ~,
&K &= \frac{k}{k'} ~.
 \label{rela}
\end{align}
Utilizing the coordinate transformation
\begin{align}
\tilde r &=  \frac{1}{K} \sqrt{r^2 - a^2} ~,&\tilde t &= K t ~,
&\tilde \theta &= K \theta ~, 
\end{align}
we can indeed rewrite the above metric into the form of \eqref{AdS3}
with the periodicity $\tilde \theta \sim \tilde \theta + 2\pi/k'$.
The boundary behavior of the metric in the form \eqref{AOS} does not
depend on the parameter $a(k')$, therefore we can express
all the orbifolds $AdS_3/{\mathbb Z}_{k'}$ $(k' < k)$ with
the same boundary condition as for $AdS_3/{\mathbb Z}_{k}$.
Notice that $k'$ should be odd since only the case with odd 
$k'$ is consistent with the anti-periodic conditions 
for fermions at the AdS boundary.

An advantage to embed into an AdS space is that the mass of
geometry is well-defined in an asymptotically AdS space.
Utilizing this fact we can analyze the stability of geometries
by comparing the masses of geometry.
Here we follow the methods developed in \cite{BK}.
For an asymptotically AdS space we can 
expand the metric for large $r$ as
\begin{align}
 d s^2 = \frac{l^2}{r^2} dr^2 + 
  \frac{r^2}{l^2} ( - dt^2 + l^2 d \theta^2 ) 
         + \delta g_{\mu \nu} dx^\mu dx^\nu ~,
         \label{asym}
\end{align}
where $\delta g_{\mu \nu}$ contains the lower powers of $r$.
Then the mass of geometry can be computed by using the formula
\cite{BK}
\begin{align}
 M = \frac{1}{8 \pi G_3} \int_0^{2 \pi /k} d \theta
  \left( \frac{r^4}{2 l^4} \delta g_{rr} + \frac{1}{l^2}
  \delta g_{\theta \theta}
     - \frac{r}{2 l^2} \partial_r \delta g_{\theta \theta} \right)
\end{align}
with the three dimensional Newton constant $G_3$. 
We find from the metric \eqref{AOS}
\begin{align}
 \delta g_{rr} &= - \frac{l^4}{r^4} \left( 1 - \frac{a^2}{l^2} \right) ~,
 &\delta g_{tt} &= - 1 ~,
 &\delta g_{\theta \theta } &= - a^2 ~,
\end{align}
thus the mass of the geometry \eqref{AOS} is given by
\begin{align}
 M = - \frac{1}{8 k G_3} \left( 1 + \frac{a^2}{l^2} \right) 
    = - \frac{k}{8 {k'}^2 G_3}  ~.
    \label{MAOS}
\end{align}
{}From this mass formula,
we can show that the mass of
geometry is largest for the original geometry
with $k' = k$ and becomes smaller as we decrease $k'$.
The final geometry should be given by $AdS_3$ with $k' =1$,
which is stable since it has the smallest
mass and no localized tachyon.
Furthermore, the supersymmetry is recovered in the final geometry.


\subsection{Gravity solution describing the decay of $AdS_3/{\mathbb Z}_k$}
\label{gravity}

In the previous subsection, we have conjectured that the
localized tachyon condensation leads to the decay of 
the orbifold $AdS_3/{\mathbb Z}_k$ into 
$AdS_3/{\mathbb Z}_{k'}$ $(k' < k)$ with a smaller 
deficit angle.
The dynamical process may be given as follows.
Tachyons localized at the fixed point could roll down
the potential and reach to minima. 
The energy due to the tachyon condensation would be 
carried out by a dilaton pulse from the center to the
boundary of the AdS orbifold.
The dilaton pulse can serve as a moving domain wall, and
the geometry decays into $AdS_3/{\mathbb Z}_{k'}$ $(k' < k)$
after the pulse passed away.
For $R_t \times {\mathbb C}/{\mathbb Z}_k$ 
this scenario was conjectured in \cite{APS} and 
the exact gravity solution was found in \cite{GH,Headrick}.

It is well known that it is difficult to analyze the condensation of 
closed string tachyon in general, since the condensation changes the 
background itself and we do not know how to deal with this case.
An advantage to localize the tachyon is that the effects of tachyon 
condensation are confined in a stringy regime, and hence
we can safely use the classical gravity description to 
describe the decay of the AdS orbifold for later time.
As mentioned above we assume that the effect of localized tachyon 
induces a dilaton pulse traveling from the center to the AdS boundary.
Thus now the problem is to find out the solution of graviton-dilaton 
system corresponding to the decay of AdS orbifold with a dilaton pulse.
The action we consider for graviton and dilaton is 
\begin{align}
 S = \frac{1}{16 \pi G_3}
 \int d^3 x \sqrt{- g} ( {\cal R} - 4 \partial_{\mu} \Phi 
\partial^{\mu} \Phi - 2 \Lambda ) ~,
\label{action}
\end{align}
where ${\cal R}$ is the Ricci scalar with respect to the metric
$g_{\mu \nu}$, and $\Phi$ is the dilaton field.
The determinant is denoted as $g = \det g_{\mu \nu}$, and the
Ricci tensor will be represented as ${\cal R}_{\mu \nu}$.
The negative cosmological constant is related as $\Lambda = -1/l^2$ in
eq.~(\ref{AdS3}) and we fix it as $\Lambda = -1$, i.e., $l = 1$
for a while.

From the action for graviton and dilaton, we can read off
the equations of motion for graviton as
\begin{align}
\label{Einsteineq}
 {\cal R}_{\mu \nu} - {\frac{1}{ 2}} {\cal R} g_{\mu \nu} =  
4 \left( \partial_{\mu} \Phi \partial_{\nu} \Phi - {\frac{1}{ 2}} g_{\mu \nu}  (\partial \Phi )^2 \right) + g_{\mu \nu}
\end{align}
and for dilaton as
\begin{align}
  \frac{1}{\sqrt{-g}} \partial_{\mu} \sqrt{-g} g_{\mu \nu} 
\partial_{\nu} \Phi= 0 ~.
\label{dilatoneq}
\end{align}
In order to solve the Einstein-dilaton equations, we set up an
initial configuration at an initial time $t=0$, and follow the
evolutions of metric and dilaton by solving these equations.
Since the $(t \mu)$ components of Einstein equations 
(\ref{Einsteineq}) contain only  
terms at most involving first derivative with respective to time $\partial_t$ and contain no second or higher time derivatives, 
we treat these equations as constraint equations for initial data. 
This is because these equations do not tell anything about time evolution. 
We treat the rest, spatial components of Einstein equations, which involve second order time derivatives,  
as dynamical evolution equations.

Since some components of Einstein equations are treated as constraint equations,
now the number of differential equations is smaller than that
of degrees of freedom. Therefore, we have to remove several components
of metric by utilizing the diffeomorphism gauge symmetry.
We can always choose the metric in the form of
\begin{align}
 ds^2 &= e^{2 F(t,r)} ( - dt^2 + dr^2 ) + C (t,r) ^2 d \theta ^2 ~,
 & \theta & \sim \theta + 2\pi/k ~.
 \label{form}
\end{align}
Here we have removed $\theta$-dependence of the metric by making use
of the symmetry of the system. The dilaton field is also set to
be independent of $\theta$. 
The conformal transformation of $(t,r)$ is a residual diffeomorphism 
which does not change the form of \eqref{form}, and
the residual gauge can be fixed by assigning appropriate 
boundary conditions at $r=0$ and initial configuration 
at the initial time $t=0$.

In terms of the metric form \eqref{form}, the geometries
before and after the tachyon condensation are given as follows.
The metric of the initial geometry $AdS_3/{\mathbb Z}_{k}$ is
\begin{align}
 ds^2 &= \frac{1}{\cos ^2 r} 
 ( - dt^2 + d r ^2 ) + \tan ^2 r d \theta ^2 ~,
\end{align}
which is obtained by replacing $r$ of \eqref{AdS3}
with $\tilde r$ 
by coordinate transformation $r = \tan \tilde r$ and
rewriting $\tilde r \to r$.
In this coordinate system,
the AdS boundary is located at $r=\pi/2$.
After the tachyon condensation the geometry
is proposed to be $AdS_3/{\mathbb Z}_{k'}$,
whose metric can be written as 
\begin{align}
ds^2 = {\frac{1}{\cos ^2 r}} (- d t^2 + d r^2) 
 + K^2 \tan^2 r d \theta^2
\label{diffrep}
\end{align}
with $K=k/k'$ as before.
Actually it is convenient for the later purpose to rewrite as
\begin{align}
 ds^2 &= \frac{1}{K^2 \cos ^2 (r/K)} 
 ( - dt^2 + d r ^2 ) + K^2 \tan ^2 (r/K) d \theta ^2
  \label{decay}
\end{align}
by rescaling coordinates as $t \to t/K , r \to r/K$. 
If we take $r \to 0$ limit, then the metric reduces to the one used 
in \cite{APS} for ${\mathbb C}/{\mathbb Z}_k$. As a result, 
comparison to ${\mathbb C}/{\mathbb Z}_k$ is more manifest in this 
metric, even though the radial boundary is shifted along the tachyon condensation from $r= \pi/2$ to $r= K \pi /2$.

Let us write down the explicit form of equations of motion by
using the metric \eqref{form}.
The constraint equations arise from $(tt),(tr)$ components of
Einstein equations \eqref{Einsteineq} as
\begin{align}
 &\partial_r F  \partial_r H  - (\partial_r H )^2 - \partial_r^2 H + 
 \partial_t F \partial_t H - 2 (\partial_t \Phi )^2 
 - 2 (\partial_r \Phi )^2 + e^{2 F} = 0~, \nonumber \\
 &\partial_r H (\partial_t F  - \partial_t H)
+ \partial_r F \partial_t H - \partial_t \partial_r H - 4
\partial_t \Phi \partial_r \Phi = 0 ~, \label{con}
\end{align}
which do not include second derivatives at it should be the case.
Here we have used $H(t,r) = \log C(t,r)$ such that the equations
become simpler.
Note that the $(t\theta)$ component is empty due 
to the $\theta$-independence.
We will use below these equations to set up initial configurations and
to check the reliability of our computation.
The non-trivial parts of evolution equations come from
$(rr)$ and $(\theta \theta)$ components as
\begin{align}
\label{NewEvoEQ}
 &\partial_t F \partial_t H +  \partial_r F \partial_r H  - (\partial_t H)^2 -  \partial_t^2 H 
  - 2 (\partial_t \Phi )^2 - 2 (\partial_r \Phi )^2 - e^{2 F} = 0 ~, \nonumber \\
  & ( \partial_t ^2 - \partial_r ^2 ) F
 - e^{2 F} - 2 (\partial_t \Phi )^2
  + 2 (\partial_r \Phi)^2 = 0 ~.
\end{align}
It is possible to solve these equations directly, but it might be
useful take a linear combination of Einstein equations to
make the equations simpler.
Notice that the Einstein equations can be reduced to a
simpler form in this case as
\begin{align}
 {\cal R}_{\mu \nu} + 2  g_{\mu \nu} &=  
4 \partial_{\mu} \Phi \partial_{\nu} \Phi ~.
\label{einstein}
\end{align}
We pick up $(tt) +(rr)$ and $(\theta\theta)$ components as evolution
equations 
\begin{align}
\label{EvoEQ}
 &2 \Delta F + \Delta H + \nabla H \cdot \nabla H 
  - 4 e^{2 F} + 4 \nabla \Phi \cdot \nabla \Phi = 0 ~, \nonumber \\ 
  &\Delta H + \nabla H \cdot \nabla H - 2 e^{2 F} = 0 ~,
\end{align}
where we have used
$\Delta = - \partial_t^2 + \partial_r^2$ and
$\nabla f \cdot \nabla f = - \partial_t f \partial_t f 
 + \partial_r f \partial_r f$.
In particular, there is no dependence of dilaton in the 
$(\theta\theta)$ component.
In this notation, the equation of motion for dilaton is
written as
\begin{align}  
  &\Delta \Phi + \nabla H \cdot \nabla \Phi = 0 ~.
  \label{dilatoneom}
\end{align}
In the following we will try to solve the three evolution
equations \eqref{EvoEQ} and \eqref{dilatoneom} for three
unknowns $F(t,r),C(t,r),\Phi(t,r)$. In fact, this is 
equivalent to solve \eqref{NewEvoEQ} and \eqref{dilatoneom}
since we have just picked up a specific linear combination.

In order to solve the evolution equations we have to set up
boundary conditions%
\footnote{Boundary conditions at the AdS boundary are tricky 
since the radial boundary shifts as tachyon condenses.
In spite of this fact, we set the Dirichlet boundary
conditions for $F,C,\Phi$ at $r=\pi/2$.
This choice is reliable only when we follow the evolution
before the dilaton pulse reaches to the boundary as below.}
at the center $r=0$ and an initial configuration at $t=0$.
At $r=0$ we set $C=0$ since the cycle of $\theta$ should
shrink at $r=0$. Then the regularity of \eqref{dilatoneom}
requires the Neumann boundary condition as 
$\partial_r \Phi = 0$ at $r=0$.
We also assign $\partial_r F = 0$ at $r=0$, which follows
the regularity of \eqref{NewEvoEQ}. This condition should
be related to the regularity of \eqref{EvoEQ} since we have 
just picked up a linear combination.
Now that we are trying to solve second order differential equations
for three unknowns, we should assign 6 initial conditions
for $F,C,\Phi$ and $\partial_t F, \partial_t C , \partial_t \Phi$ at $t=0$.
At the initial time we have argued that the tachyon condensation makes
a dilaton pulse, which should be determined from the string theory 
computation in principle. Since it is a rather hard task, we simply 
assume that the localized tachyon induces a static dilaton pulse with
the Gaussian form as
\begin{align}
 \Phi(t=0,r) &= \Phi_0 \exp ( - r^2 / \Delta) ~,
 &\partial_t \Phi (t=0,r ) &= 0 ~. \label{iniP}
\end{align}
The normalisation $\Phi_0$ and the width $\sqrt \Delta$ of the pulse
should be related to the localized tachyon condensation and 
therefore to the decay process.
We also assume that the decay starts from a static
configuration and hence we set $\partial_t F = \partial_t C = 0$
at $t=0$.
The other intimal conditions are for $F$ and $C$ at $t=0$.
Due to the assumption of static initial configuration, 
the second equation of \eqref{con} vanishes.
Therefore, once we fix one of the initial conditions, then the other
is determined from the first constraint equation.
We fix it from the flat space limit.
Near $r=0$ we can neglect the cosmological constant,
thus the change of metric can be close to the one in \cite{GH,Headrick}. 
They fix $C=r$, independent of dilaton pulse,
with the help of the residual diffeomorphism, thus we may set 
$C= \tan r$ at $t=0$. Then $F$ at $t=0$ is determined by solving 
the first constraint equation.
Because of this choice of initial configuration, we expect that 
the decay of $AdS_3/{\mathbb Z}_k$ is closed to the one of
${\mathbb C}/{\mathbb Z}_k$ at least near $r=0$.
In particular, the metric of the final geometry should be given 
as in \eqref{decay}.

\begin{figure}[htbp]
 \begin{minipage}{0.49\hsize}
  \begin{center}
   \includegraphics[width=78mm]{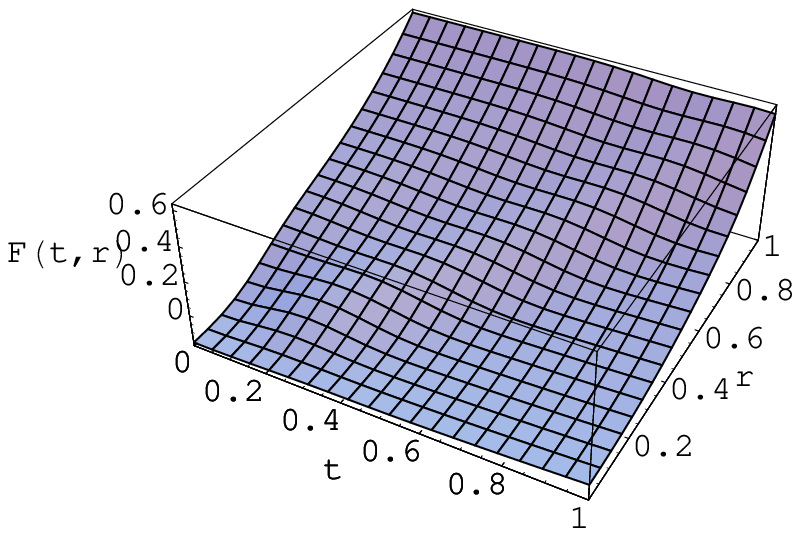}
  \end{center}
 \end{minipage}
 \begin{minipage}{0.49\hsize}
  \begin{center}
   \includegraphics[width=47mm]{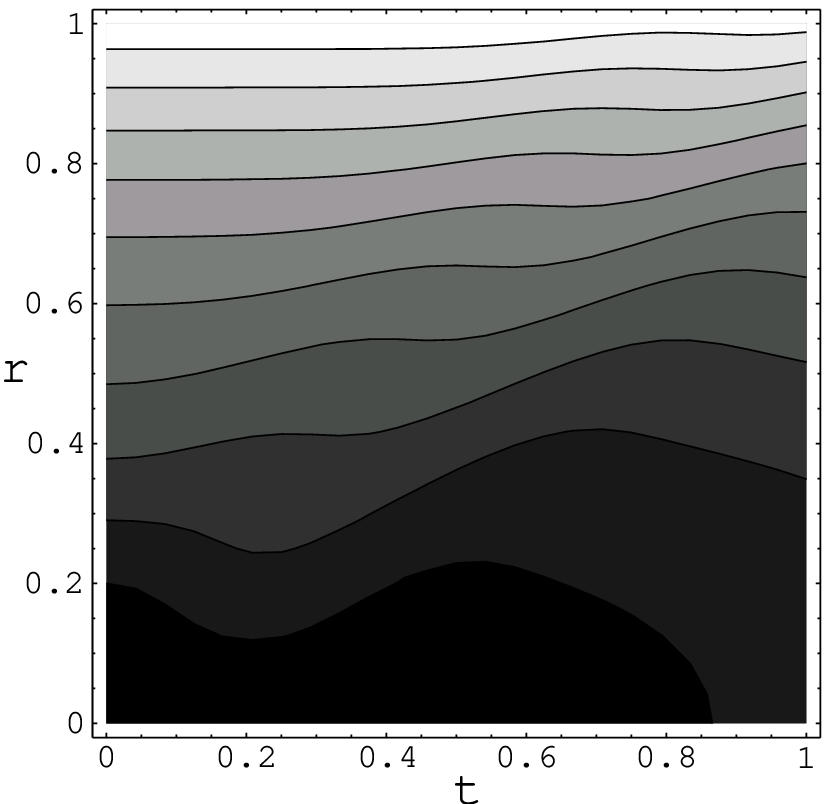}
  \end{center}
 \end{minipage} \\
 \begin{minipage}{0.49\hsize}
  \begin{center}
   \includegraphics[width=78mm]{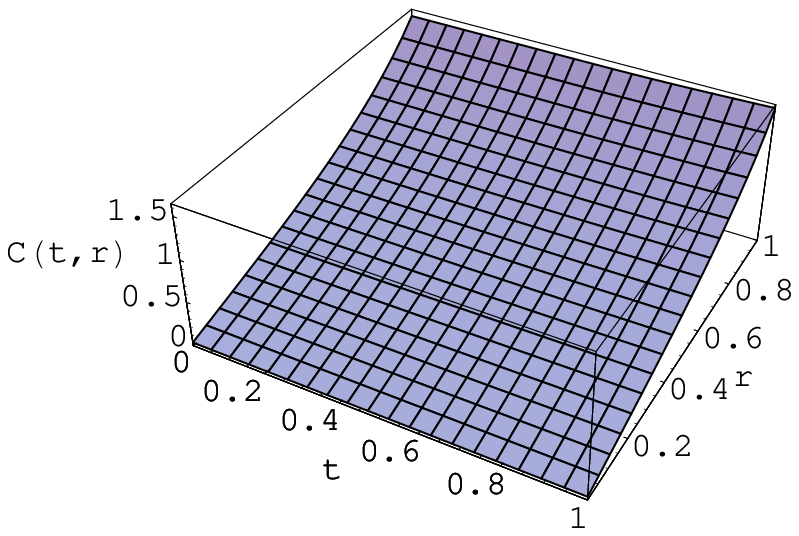}
  \end{center}
 \end{minipage}
 \begin{minipage}{0.49\hsize}
  \begin{center}
   \includegraphics[width=47mm]{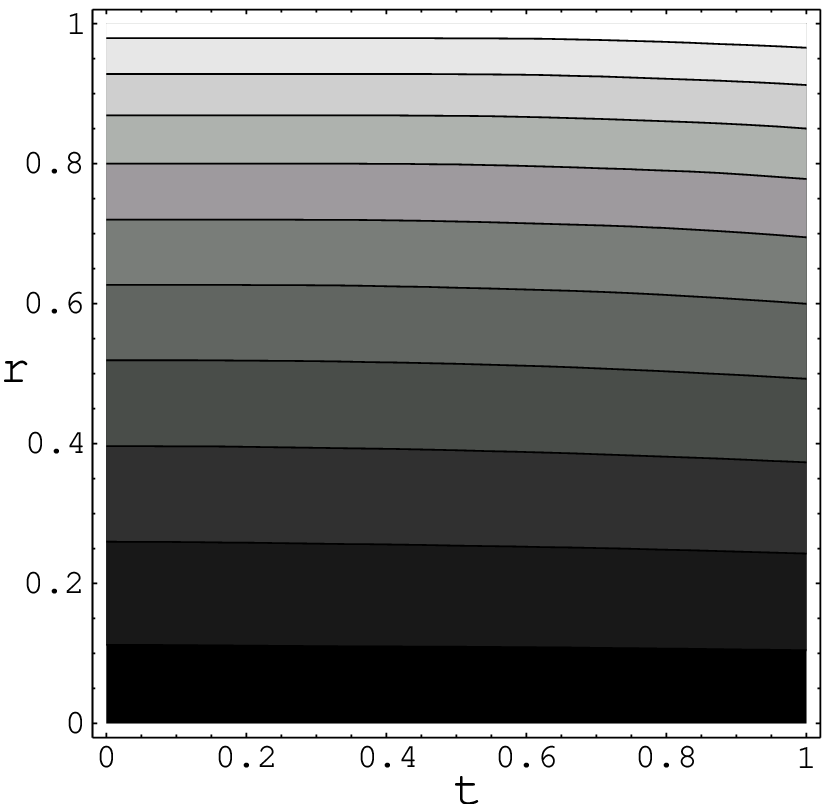}
  \end{center}
 \end{minipage} \\
 \begin{minipage}{0.49\hsize}
  \begin{center}
   \includegraphics[width=78mm]{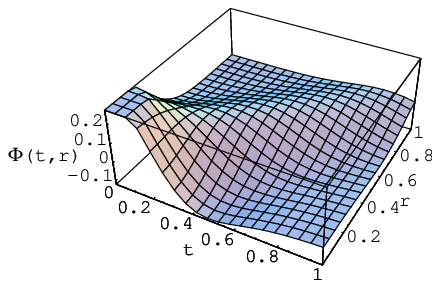}
  \end{center}
 \end{minipage} 
 \begin{minipage}{0.49\hsize}
  \begin{center}
   \includegraphics[width=47mm]{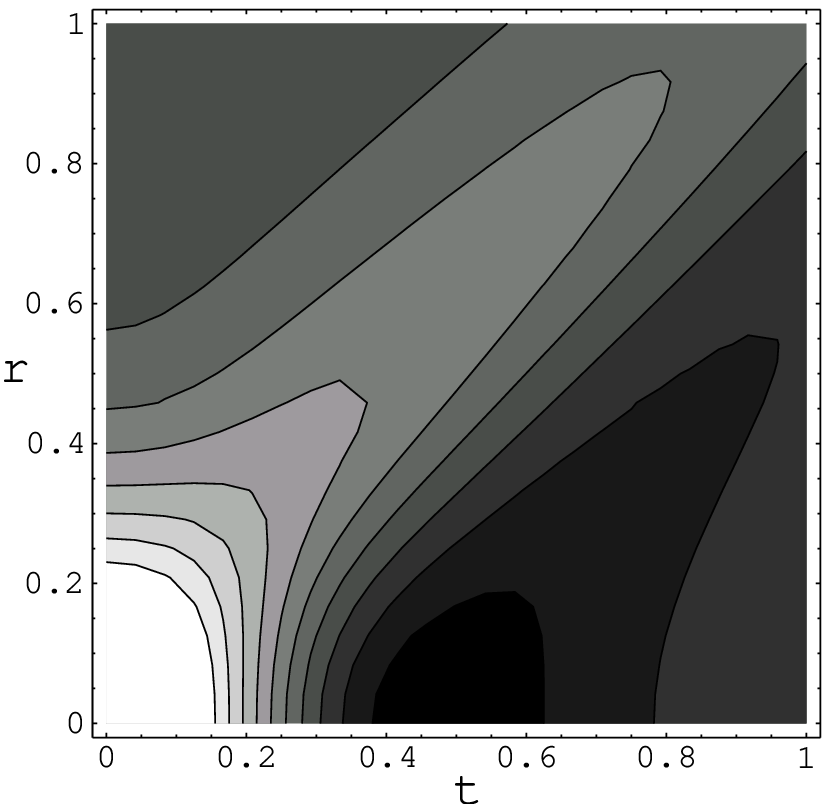}
  \end{center}
 \end{minipage}
  \caption{\it The 3D plots of the solutions $F,C,\Phi$ to the
evolution equations \eqref{EvoEQ} and \eqref{dilatoneom}.
 The right hand sides are the contour plots.
 In the numerical computation we have set 
 $\Phi_0 = 0.4$ and $\Delta = 0.1$ in \eqref{iniP}. 
 The function $C$ changes very little during the process, and
 the dilaton $\Phi$ travels as in the static AdS orbifold.
 The function $F$ decreases after the dilaton pulse passed away,
 which is consistent with the conjecture of localized tachyon
 condensation.
}
  \label{ads}
\end{figure}

Right now we have sufficient boundary conditions to solve the
three evolution equations \eqref{EvoEQ} and \eqref{dilatoneom}.
Unfortunately we cannot find analytic solutions to these equations,
therefore we try to solve them in a numerical way.
The result is summarized in fig.~\ref{ads}, 
and the constraint equations \eqref{con} are checked numerically. 
We interpret the result as follows. 
The function $C$ changes very little during
the decay process, which is consistent with the assumption of
initial condition.
Since the equation of motion for dilaton depends only on $C$ but not
on $F$, the dilaton pulse is almost the same as in the static
AdS orbifold. 
The most important information should be read off from the behavior
of $F$. We can see that the value of $F$ decreases after the dilaton
pulse passed away.
This is consistent with the expectation of \eqref{decay}
that $F$ changes as $F \to F - \log K$ for small $r$ by the 
tachyon condensation.
In this way, at least until dilaton pulse reaches the boundary, 
we have checked numerically that the evolution of tachyon condensation 
is consistent with the proposal that $AdS_3/{\mathbb Z}_k$ decays 
into $AdS_3/{\mathbb Z}_{k'}$ with $k' < k$.


\subsection{Dual gauge theory description}
\label{gauge3}

The AdS/CFT correspondence relevant for this case may be 
deduced from the near horizon limit of D1/D5 system 
\cite{Maldacena}.
We wrap $N_5$ D5-branes over a small $T^4$, which 
gives string-like objects in (1+5) dimensional
space-time. We put $N_1$ D1-branes on the top of
the string-like objects, then the near horizon limit
of the D1/D5 system gives rise to closed superstrings 
on $AdS_3 \times S^3 \, (\times T^4)$.
The dual gauge theory could be described by the low energy 
effective action on the worldvolume of the D1/D5 system.
For the dual of the orbifold $AdS_3/{\mathbb Z}_k$, 
we should consider the orbifold of the worldvolume
theory defined on its boundary $R_t \times S^1/{\mathbb Z}_k$.

In this subsection we restrict ourselves to large $N_1,N_5$
and zero 't Hooft coupling limit. 
The radius of $S^1$ is related to the AdS radius as $R=l$,
which may be read from the asymptotic behavior of the metric
\eqref{asym}.
In order to make the 't Hooft coupling small, we have
to consider the case with a small radius $R$.
If we take the IR limit or the large radius limit, then the gauge 
theory description is not valid anymore due to the large 't Hooft 
coupling, and conformal field theory description should be adopted.

Two ends of open strings can be attached to either of 
D1-brane or D5-brane. {}From the open strings between the same
D-brane, we obtain $U(N_1)$ and $U(N_5)$ gauge fields in the 
low energy limit. In (1+1) dimension, almost all the degrees of
freedom can be gauged away and only the zero modes are left.
The zero modes of gauge fields induce the holonomies $V_1$ for $U(N_1)$ 
and $V_5$ for $U(N_5)$ along $S^1/{\mathbb Z}_k$ spatial cycle, 
and the choice of holonomies labels the vacua of the theory.
For the open strings between D1 and D5-branes,
we can find from some computations that the low energy 
spectrum includes 4 bi-fundamental scalers and 4 bi-fundamental 
fermions with respect to the $U(N_1) \times U(N_5)$ gauge symmetry
(see, e.g., \cite{CM}).
We use odd $k$ and assign anti-periodic boundary conditions
for fermions along the spatial cycle.

In order to classify the possible vacua of the orbifold gauge 
theory, we have to find out which holonomies could be taken.
If we consider the gauge theory on the covering space, then
only the trivial holonomies along the spatial cycle are allowed. 
This can be understood as follows.
Consider a two dimensional $U(N)$ gauge theory on a 2-torus
$T^2$ and a holonomy matrix along its {\it thermal} cycle. 
Then for large $N$ limit the eigenvalues of holonomy matrix are 
uniformly distributed at low temperature and collapsed at high temperature.
In particular, in the infinite temperature limit, the density of eigenvalue 
becomes delta-functional. For example, see \cite{T2}.
Thus holonomies along {\it thermal} cycle are trivial in the 
infinite temperature limit. But from the modular invariance, this means that   
holonomies along the {\it spatial} cycle are trivial in the zero temperature limit. Therefore, we have the conditions $V_1^k = 1$ and $V_5^k = 1$ 
for the orbifold gauge theory
as in the case of $R_t \times S^3/{\mathbb Z}_k$ \cite{LM}.
Utilizing the gauge symmetry we can set $V_1$ and $V_5$ in the form
of
\begin{align}
 &{\rm diag} (1 , \cdots 1, \omega , \cdots , \omega , \cdots 
            \omega^{k-1} , \cdots , \omega^{k-1}) ~,
 &\omega &= \exp \frac{2 \pi i}{k} ~. 
\end{align}
In other words, the vacua are labeled by the $2k$ integer numbers
$(n^1_0 , \cdots , n^1_{k-1})$ and $(n^5_0 , \cdots , n^5_{k-1})$,
where the numbers of $\omega^I$ are denoted as $n^1_I$ and $n^5_I$ 
with $\sum_I n^1_I = N_1$ and $\sum_I n^5_I = N_5$.

Let us examine the spectrum of this orbifold gauge theory.
Due to the existence of non-trivial holonomies, the gauge
symmetry is broken to $\prod_I U(n^1_I) \times \prod_J U(n^5_J)$
and the states are in the bi-fundamental representation of 
this broken gauge symmetry.
First we consider a scalar in the $(n^1_I , \bar n^5_J)$
bi-fundamental representation.
The scalar can be expanded by a plane wave as
$\Phi^{(I,J)}_p \sim e^{i p \theta}$,
and the orbifold action $g$ yields
\begin{align}
g \cdot \Phi^{(I,J)}_p =  
 e^{\frac{2 \pi i p}{k}} \omega^{I-J} \Phi^{(I,J)}_p 
 = e^{\frac{2 \pi i}{k}(p + I - J)} \Phi^{(I,J)}_p ~.
\end{align}
Notice that the phase factor is shifted by the effect of
holonomies.
The orbifold invariant states can be obtained by 
summing over all the images of the orbifold action.
Thus the projection operator is given by 
$\Gamma = \sum_{I=1}^{k}g^I $, and the action of this operator
restricts the modes to $p = k n + J - I$ with $n \in {\mathbb Z}$.
The spectrum of fermion in the $(n^1_I , \bar n^5_J)$
bi-fundamental representation can be obtained in the same way.
Here we should remember that $k$ is an odd integer and 
the anti-periodic boundary condition is assigned. 
Thus the orbifold action becomes
\begin{align}
g \cdot \Psi^{(I,J)}_p = 
e^{\frac{2 \pi i}{k}(p + I - J +  k/2)} \Psi^{(I,J)}_p ~,
\end{align}
where the shift of $k/2$ arises from the anti-periodic boundary
condition. The projection into the orbifold invariant subspace
leads to the restriction $p = k (n + 1/2) + J - I$ with 
$n \in {\mathbb Z}$.

Now that we know the spectrum of the orbifold theory for
arbitrary holonomy matrices, we can compute the Casimir energy,
which is known to be dual to the mass of dual geometry.
The Casimir energy is given by
\begin{align}
V_0 &= \frac{1}{2} \sum_E (-1)^F E n_E ~,
&E &= \frac{|p|}{R} ~,
\end{align}
where $F$ denotes the fermion number and 
$n_E$ represents the number of states with
energy $E$.
For a scalar or a fermion in the $(n^1_I , \bar n^5_J)$ 
bi-fundamental representation 
the number of modes is $n^1_I n^5_J$ and the energy is 
$E =  |k n  + I - J|/R$ for a scalar and 
$E =  |k (n + 1/2)  + I - J|/R$ for a fermion.
Since the orbifold gauge theory includes 4 scalars and
4 fermions, we obtain
\begin{align}
 V_0 = \sum_{I,J} \frac{ n^1_I n^5_J}{2 R} 
 \left( 4 \sum_{n= - \infty}^{\infty} |k n  + I - J| - 
   4 \sum_{n= - \infty}^{\infty} 
 |k (n+1/2)  + I - J | \right) ~.
\end{align}
Using this formula, we can compute the Casimir energy
for each vacuum with generic holonomies.

We would like to find out the vacuum dual to $AdS_3/{\mathbb Z}_{k'}$
with $k' \leq k$.
The orbifold $AdS_3/{\mathbb Z}_{k'}$ has ${\mathbb Z}_{k'}$ symmetry,
thus the holonomy matrices should respect this discrete symmetry.
If we restrict ourselves to the case with integer $K = k/k'$,
then we can choose $n^1_{m K} = N_1/k'$ and $n^5_{m K} = N_5/k'$
with $m = 0,1,\cdots k' - 1$ 
and zero for others.%
\footnote{
Here we have assumed that $N_1$ and $N_5$
can be divided by $k'$, though the precise value is not relevant
for large $N_1,N_5$.}
There might be other choices of holonomies 
respecting the symmetry, but we can show that this choice gives 
the smallest Casimir energy among them.
In fact, the Casimir energy in this case is computed as
\begin{align}
 V_0 = \frac{N_1 N_5}{R {k'}^2 } k'
 \sum_{I=1}^{k'} \left( 4 \sum_{n=1}^{\infty} |kn + K I| - 
   4 \sum_{n=1}^{\infty} |k( n +1/2) + K I | \right)
   = - \frac{c k}{12 {k'}^2 R} 
\end{align}
with $c = 6 N_1 N_5$.
In order to obtain this, it is useful to use the formula
\begin{align}
 \sum_{n=1}^{\infty} ( n - \theta ) = 
 \frac{1}{24} - \frac{1}{8} (2 \theta - 1 )^2 ~.
\end{align}
Using the relation $c= 3 l/(2 G_3)$ (see, e.g., \cite{BK}),
the Casimir energy exactly matches the mass of $AdS_3/{\mathbb Z}_{k'}$ 
\eqref{MAOS}. 
Notice that we obtain the exact match contrary to the 
$AdS_5/{\mathbb Z}_k$ case \cite{Hikida}.
This could be another example showing that $AdS_3$ cases are more
stable under quantum corrections than $AdS_5$ cases,
which is known to occur in many contexts.

One may ask what would happen for generic $k'$ with
non-integer $K = k/k'$.
The answer depends on whether we deal with infinite or
finite $N_1,N_5$. For infinitely large $N_1,N_5$, we may be able to construct
a vacuum arbitrary close to the dual of each geometry.
For finite $N_1,N_5$, we have a finite number of vacua, 
thus not all of the classical geometries have their dual vacua.
If we include quantum conditions to the gravity side, then only
the geometries with dual gauge theory vacua may be allowed.%
\footnote{See, for example, \cite{LM}. In their case 
the quantization of flux restricts the number of allowed geometry 
and leads to one-to-one correspondence between geometries and 
gauge theory vacua.}

We conclude this subsection as follows. 
We may start from the vacuum dual to the $AdS_3/{\mathbb Z}_k$, 
which is labeled by the holonomies $n_I^1 = N_1/k$ and 
$n_I^5 = N_5/k$ for all $I$. This vacuum is only meta-stable because
other vacua have smaller Casimir energies.
The vacuum decays non-perturbatively into another 
vacuum dual to $AdS_3/{\mathbb Z}_{k'}$ with a smaller $k'$,
and finally ends up with the trivial 
vacuum with $n_0^1 = N_1$ and $n_0^5 = N_5$, which is 
dual to pure $AdS_3$. The vacuum transition will be 
discussed in the next section for $AdS_5$ case.

\section{Decay of $AdS_5/{\mathbb Z}_k$}
\label{ads5}

As we saw in the previous section, the localized tachyon condensation
on $AdS_3/{\mathbb Z}_k$ leads to the decay of geometry
in a quite analogous way to the decay of 
${\mathbb C}/{\mathbb Z}_k$.
However, the localized tachyon condensation on $AdS_5/{\mathbb Z}_k$ 
is quite different as discussed in \cite{Hikida}. 
In fact, $AdS_5/{\mathbb Z}_k$ does not decay into $AdS_5$ or 
the other orbifold of $AdS_5$, because 
 the boundary of $AdS_5/{\mathbb Z}_k$, i.e., $R_t \times S^3/{\mathbb Z}_k$,
cannot be the boundary of $AdS_5$ or the AdS orbifold 
$AdS_5/{\mathbb Z}_{k'}$ with $k ' \neq k$. 
The final geometry after the tachyon condensation was proposed
in \cite{CM1,CM2} and called as Eguchi-Hanson soliton.

The dual gauge theory description can be given by the 
${\mathbb Z}_k$ orbifold of ${\cal N}=4$ super Yang-Mills on 
$R_t \times S^3$.
In particular, the Casimir energies of various vacua 
were computed in \cite{Hikida}, and it was found that the 
Casimir energies reproduce the masses of dual geometries quite well.
Similar results were obtained in \cite{NT} in 
slightly different configurations.
In the next subsection, we review the work of \cite{Hikida},
which discuss the fate of localized tachyon condensation 
on $AdS_5/{\mathbb Z}_k$ and its gauge theory description.
This subsection is for the preparation of subsection \ref{gauge5},
where the transition between different vacua is discussed.
The transition is described by an instanton of the orbifold gauge 
theory on $R_t \times S^3/{\mathbb Z}_k$.
We construct instanton solutions by making use of the known 
instantons for the gauge theory on $R_t \times S^2$ \cite{PhD,Lin}.


\subsection{Review of final geometry and dual gauge theory description}
\label{final5}

We consider type IIB superstring theory on $AdS_5 \times S^5$
and construct the orbifold theory with tachyonic modes at the fixed
point. In the global coordinates the metric of $AdS_5$ is given by
\begin{align}
 ds^2 &= g(r) dt^2 + \frac{dr^2}{g(r)}
      + r^2 d \Omega_3 ~, 
 &g(r) &= r^2  + 1 ~,
 \label{AdS5}
\end{align}
where the AdS radius is set to be one and the metric of
boundary geometry is 
\begin{align}
  d \Omega_3 = \frac{1}{4}    
  \left[ ( d \chi + \cos \theta d \phi )^2 + d \theta ^2
            + \sin ^2 \theta d \phi ^2 \right] ~.
            \label{bm}
\end{align}
The variables run $0 \leq \theta \leq \pi$, $0 \leq \phi \leq 2 \pi$,
and $0 \leq \chi \leq 4 \pi$.
The identification in the orbifold theory is performed by the
shift along the $\chi$-cycle as $\chi \sim \chi + 4 \pi/k$.
Then the orbifold action yields a fixed point at $r=0$, and
the system has tachyonic modes localized at the fixed point,
if we use even%
\footnote{One may ask why $k$ should be even contrary to the
$AdS_3$ case with odd $k$. This is related to the
topology of boundary geometry. 
The boundary of $AdS_3$ is given by $S^1$, and the cycle can be 
pinched off at the center of $AdS_3$ if we assign the anti-periodic boundary 
condition for fermions. This leads to the condition of odd $k$ 
for the orbifold theory. On the other hand, the boundary of $AdS_5$
is $S^3$, and there is no cycle which we can go around. 
For this reason
we can assign anti-periodic boundary conditions only for even $k$
such that fermions do not receive a phase factor when going around
$k$ times the cycle of $S^3/{\mathbb Z}_k$.
}
$k$ and assign the anti-periodic boundary condition
for fermions along the $\chi$-cycle.

The geometry after the localized tachyon condensation is 
proposed in \cite{CM1,CM2}, 
where they called the geometry as Eguchi-Hanson soliton.
The metric is given by
\begin{align}
 ds^2 = g(r) dt^2 + \frac{dr^2}{g(r)f(r)}
      + \frac{r^2}{4}
       \left[ f(r) ( d \chi + \cos \theta d \phi )^2 + d \theta ^2
            + \sin ^2 \theta d \phi ^2 \right]
            \label{EH}
\end{align}
with
\begin{align}
  g(r) &= r^2  + 1 ~,
&f(r) &= 1 - \frac{a^4}{r^4} ~, 
&a^2 &= \left(\frac{ k^2}{4} - 1\right) ~,
\label{defa}
\end{align}
where $k > 2$ such that $a^2 > 0$.
The relation between $a$ and $k$ is fixed by assuming the
regularity of the geometry at $r=a$, and due to the lack of 
fixed point the Eguchi-Hanson soliton does not have 
localized tachyons.%
\footnote{It might be interesting to use a generic $a$ to 
construct other geometry with an orbifold singularity at $r=a$.
It may serve as an intermediate geometry.}
The region of $r < a$ is removed in the 
Eguchi-Hanson soliton, and this region might be interpreted as 
the tachyon state, where the tachyonic modes have non-trivial 
expectation values \cite{Horowitz,HS}.
We can check by taking large $r$ limit that this geometry has 
the same boundary geometry as that of $AdS_5/{\mathbb Z}_k$, 
whose metric is given by \eqref{bm}.

We can discuss the stability of background by comparing the 
masses of geometry addition to examining the existence of
tachyonic modes.
The mass of the AdS orbifold $AdS_5/{\mathbb Z}_k$ is just 
$1/k$ times that of $AdS_5$, thus it is given by
\begin{align}
M = \frac{3 \pi}{32 k G_5} ~.
\label{massAdS}
\end{align}
The mass of Eguchi-Hanson soliton was computed in \cite{CM1,CM2}
by adopting the same method in subsection \ref{final3} as
\begin{align}
 M = - \frac{\pi ( k^4 - 8k^2 + 4)}{128 k G_5} ~.
 \label{massEH}
\end{align} 
We can see that  the mass of Eguchi-Hanson soliton is smaller than  
that of $AdS_5/{\mathbb Z}_k$, and hence the Eguchi-Hanson soliton
can be thought as a final geometry.

The gauge theory dual to superstring theory on 
$AdS_5/{\mathbb Z_k} \times S^5$ is given by
${\cal N} = 4$ $U(N)$ super Yang-Mills theory on 
$R_t \times S^3/{\mathbb Z_k}$ \cite{LM}.
The radius of $S^3$ is set to be one and $N$ is 
taken very large. We use the metric of 
$R_t \times S^3/{\mathbb Z_k}$ as
\begin{align}
 ds^2 = - dt^2 + \frac{1}{4}
   \left[ ( d \chi + \cos \theta d \phi )^2 + d \theta ^2 
    + \sin ^2 \theta d \phi ^2 \right] ~,
\end{align}
where the theory is divided by the shift of $2 \pi /k$ 
along the $\chi$-cycle.
Originally there is no non-trivial cycle in the covering space $S^3$,
but the orbifold procedure leads to a non-trivial cycle with 
$\pi_1 (S^3/{\mathbb Z}_k) = {\mathbb Z}_k$.
Along the cycle, we can assign a holonomy matrix 
$V = P \exp (- i g_{YM} \oint {\cal A}_{\chi})$ subject to $V^k = 1$
as in the $AdS_3$ case.
The holonomy matrix can be set as
\begin{align}
V &= {\rm diag}
 (1,\cdots,1,\omega,\cdots,\omega,\cdots,\omega^{k-1},\cdots\omega^{k-1}) ~,
 &\omega &= \exp \frac{2 \pi i}{k} 
 \label{hol5}
\end{align}
with the help of $U(N)$ gauge symmetry.
Therefore, the vacua are labeled by $k$ integers $(n_0,\cdots,n_{k-1})$
with $\sum_I n_I = N$, where $n_I$ represents the number of $\omega^I$.
Two specific vacua among them are important for us.
One is the vacuum with the ${\mathbb Z}_k$ symmetric holonomy 
$n_I = N/k$ for all $I$, which is dual to 
$AdS_5/{\mathbb Z}_k$. The other is the vacuum with the trivial holonomy
$n_0 = N$, which is dual to the Eguchi-Hanson soliton \eqref{EH}.

In \cite{Hikida} the spectrum of the orbifold gauge theory with 
the holonomy matrix was obtained and the Casimir energy for the
vacuum was computed at the one loop level. 
For the ${\mathbb Z}_k$ symmetric holonomy $n_I = N/k$, 
the Casimir energy is given by
\begin{align}
 V_0 &= N^2 \frac{3}{16 k} ~.
\end{align}
With the relation $N^2 = \pi/(2 G_5)$ we can see that the Casimir energy
exactly reproduces the mass of $AdS_5/{\mathbb Z}_k$ \eqref{massAdS}.
For the trivial holonomy $n_0 = N$, the Casimir energy is
\begin{align}
 V_0 = - N^2 \left( \frac{k^3}{48} - \frac{k}{12} - \frac{3}{16 k} \right) ~,
\end{align}
which is roughly $4/3$ times the mass of the Eguchi-Hanson 
soliton \eqref{massEH}. This is a remarkable result since we have observed a 
quantitative correspondence between the results in small and 
large 't Hooft coupling limits.
We can show that the Casimir energy for $n_0 = N$ is smallest among
the ones for every holonomies \cite{Hikida}, and in this way we 
may say that the Eguchi-Hanson soliton is really the final 
geometry after the decay of $AdS_5/{\mathbb Z}_k$.


\subsection{Gauge theory instanton}
\label{gauge5}

We have observed that the localized tachyon condensation 
deforms the background geometry from AdS orbifold into 
another more stable geometry. In particular, the dynamics of
the geometry transition for $AdS_3$ case have been analyzed by 
constructing a numerical gravity solution describing 
the decay of $AdS_3/{\mathbb Z}_k$ in subsection \ref{gravity}. 
In this subsection, we would like to discuss the dynamics of 
the transition from the viewpoint of the dual gauge theory.
Each geometry corresponds to a vacuum of dual gauge theory, 
thus the transition of geometry should be described by 
the transition between different vacua, i.e., the instanton
interpolating vacua.
We focus on the orbifold gauge theory on 
$R_t \times S^3/{\mathbb Z}_k$ since we have a lot of
knowledge about instantons in four dimension.

We would like to construct instantons which interpolate
vacua at $\tau = - \infty$ and other vacua at 
$\tau = \infty$ with the Euclidean time $\tau = i t$.
We only analyze in the semi-classical limit, 
where all the vacua are degenerated, and in this limit
it is enough to excite only the gauge field.
For this reason we consider $SU(N)$ pure Yang-Mills theory,
whose action is given by
\begin{align}
 S = \frac{1}{4} \int d ^4 x \sqrt{g_4} 
  {\cal F}_{\mu \nu} { \cal F}^{\mu \nu} ~,
\label{actionS3}
\end{align}
where the field strength is defined as
\begin{align}
 {\cal F}_{\mu \nu} = \partial_{\mu} {\cal A}_{\nu}
  - \partial_{\nu} {\cal A}_{\mu} 
  + i g_{\rm YM} [{\cal A}_{\mu} , {\cal A}_{\nu}] ~.
\end{align}
We denote the Yang-Mills coupling constant as $g_{YM}$,
which is assumed to be very small.
The gauge theory is defined on $R_t \times S^3/{\mathbb Z}_k$, 
whose metric is given by
\begin{align}
 ds^2 = d \tau^2 + \frac{1}{4}
   \left[ ( d \chi + \cos \theta d \phi )^2 + d \theta ^2 
    + \sin ^2 \theta d \phi ^2 \right] 
\end{align}
with $0 \leq \theta \leq \pi$, $0 \leq \phi \leq 2 \pi$,
and $0 \leq \chi \leq 4 \pi/k$ as before.
In particular, the measure is given by 
$d^4x \sqrt{g_4}= \frac{1}{8} \sin \theta d \tau d \chi d \theta d\phi$.

In order to obtain instanton solutions, it is useful to
rewrite the above action as
\begin{align}
 S = \frac{1}{8} \int d ^4 x \sqrt{g_4}
   \left[ ({\cal F}_{\mu \nu} \mp * {\cal F}_{\mu \nu})
          ({\cal F}^{\mu \nu} \mp * {\cal F}^{\mu \nu})
           \pm 2  {\cal F}_{\mu \nu} * {\cal F}^{\mu \nu}
           \right]
           \label{actionS3_2}
\end{align}
as usual. The Hodge dual is given by
\begin{align}
  * {\cal F}_{\mu \nu} &= \frac{\sqrt{g_4}}{2 !}
   \epsilon_{\mu \nu \rho \sigma} {\cal F}^{\rho \sigma} ~,
 & * {\cal F}^{\mu \nu} &= \frac{1}{2 ! \sqrt{g_4}}
   \epsilon^{\mu \nu \rho \sigma} {\cal F}_{\rho \sigma}
\end{align}
in a curved space.
The second term of \eqref{actionS3_2} corresponds to
a topological contribution.
Within the same topological sector, the minimum of the action
is given by the solutions to the (anti-)self-dual equation
of field strength
\begin{align}
{\cal F}_{\mu \nu} = \pm * {\cal F}_{\mu \nu} ~. 
\label{sde}
\end{align}
The solutions to the equation are the (anti-)instantons
of the orbifold gauge theory.

We try to find out solutions to the (anti-)self-dual
equations. One easy guess is to utilize the 't Hooft instanton,
but this type of instantons do not interpolate the vacua of
our type.%
\footnote{This type of instantons can be constructed by
the orbifold images of the 't Hooft instantons mapped on 
$R_\tau \times S^3$.
These instantons have the topological charge ${\mathbb Z}/k$ 
and are dual to fractional instantons localized at the fixed point of $AdS$ orbifold.
In particular, the sum of all types of fractional instantons
should reproduce the bulk instanton.
}
Therefore we should look for other type of solution.
The main idea is as follows. 
Just like monopole solutions do not depend on time coordinate, 
we assume the coordinate independence along the $\chi$ direction. 
Then we can perform the dimensional reduction along the $\chi$ 
direction, and the theory is reduced to the one on $R_\tau \times S^2$.%
\footnote{The relation between gauge theories on 
$R_t \times S^3/{\mathbb Z}_k$ and on $R_t \times S^2$ was 
also discussed in \cite{ISTT}.}
Instanton solutions of the gauge theory on $R_\tau \times S^2$ were 
obtained in \cite{PhD,Lin} (see also \cite{Anders}),
thus we can obtain instantons on
$R_\tau \times S^3/{\mathbb Z}_k$ by making use of the results on
$R_\tau \times S^2$.

The dimensional reduction in this case is a little bit subtle
since $S^3$ consists of a non-trivial $S^1$ fibration over $S^2$.
Using the standard technique of Kaluza-Klein dimensional reduction, 
the gauge field
on $R_\tau \times S^2$ can be defined as \cite{LM}
\begin{align}
 {\cal A}_{\mu} d x^{\mu} = 
 A_m d x^m + \Phi (d \chi + \cos \theta d \phi )
 \label{newgauge}
\end{align}
with $m= \tau , \theta , \phi$.
After the integration over the $\chi$ direction,
we obtain the new action for the redefined gauge field as
\begin{align}
 S = \frac{4 \pi}{k} \int d ^3 x \sqrt{g_3}  
     \left[  F_{\tau \theta}^2 + 
     \frac{4}{\sin ^2 \theta} ( F_{\theta \phi} - \Phi \sin \theta )^2
     +   \frac{1}{\sin ^2 \theta} F_{\tau \phi}^2
     + D_m \Phi D^m \Phi \right] ~,
     \label{s2action}
\end{align}
where the field strength and the covariant derivative are
\begin{align}
 {F}_{mn} &= \partial_{m} {A}_{n}
  - \partial_{n} {A}_{m} 
  + i g_{\rm YM} [{A}_{m} , {A}_{n}] ~,
  &D_m \Phi &= \partial_m \Phi + i g_{YM} [A_m , \Phi ] ~.
\end{align}
The index is raised in \eqref{s2action} by
the metric of $R_{\tau} \times S^2$
\begin{align} 
 ds^2 = d \tau^2 + \frac{1}{4}
   \left[ d \theta ^2 + \sin ^2 \theta d \phi ^2 \right] ~,
\end{align}
and the measure in this case is given by
$d^3 x \sqrt{g_3} = \frac{1}{4} \sin \theta d \tau d \theta d \chi$.

The second term of \eqref{s2action} acts important roles
on the gauge theory on $R_{\tau} \times S^2$.
This term arises through the non-trivial relation 
${\cal F}_{\theta\phi} = F_{\theta\phi} - \Phi \sin \theta +
(D_\theta \Phi) \cos \theta$, where the contribution from
$(D_\theta \Phi) \cos \theta$ does not appear in the final form.
Because of the form of complete square, we can see that 
the vacuum of this gauge theory is labeled by $\Phi=f$
with the notation 
$F_{\theta \phi} d \theta d \phi = f \sin \theta d \theta d\phi$.
Through the relation \eqref{newgauge} the holonomy matrix 
$V = P \exp (- i g_{YM} \oint {\cal A}_{\chi})$ of \eqref{hol5}
is mapped to the configuration 
\begin{align}
 \Phi = f = \frac{1}{g_{\rm YM}} 
 (0,\cdots,0,1,\cdots,1,\cdots,k-1,\cdots,k-1) ~,
 \label{vacuaS2}
\end{align}
where the number of $I=0,\cdots,k-1$ is given by $n_I$
defined above.

Let us focus on the instanton case. Then the problem is now
to find out solutions to the self-dual equation \eqref{sde}
in terms of gauge field of the three dimensional theory \eqref{newgauge}. 
For $SU(2)$, the general solutions were constructed in \cite{PhD}. 
For $SU(N)$ with general $N$ it was pointed out in \cite{Lin} 
that the general solutions can be deduced from the ones in the 
plane wave matrix model \cite{BMN} obtained in \cite{YY}.
Given a solution to the self-dual equation \eqref{sde}, 
the action can be written as
\begin{align}
 S &= \frac{1}{4} \int d ^4 x \sqrt{g_4}
           {\cal F}_{\mu \nu} * {\cal F}^{\mu \nu}
       = \frac{4 \pi}{k}
     \int d \tau d \theta d \phi
      \left[ D_{\tau} \Phi( F_{\theta \phi} - \Phi \sin \theta )
       + F_{\tau \theta} D_{\phi} \Phi - F_{\tau \phi} D_\theta \Phi \right]
        \nonumber \\  &
   = \frac{2 \pi}{ k} \int d \tau d \theta d \phi \sin \theta
       D_{\tau} \Phi^2 = 
      \frac{2 \pi}{k } \int d \theta d \phi \sin \theta
       \left[ \Phi^2|_{\tau=\infty} - \Phi^2|_{\tau= -\infty} \right]
\end{align}
with the help of Bianchi identity 
$D_\tau F_{\theta \phi} +D_\theta F_{\phi \tau} 
+ D_\phi F_{\tau \theta}  = 0$ \cite{PhD}.
At the initial time $\tau = - \infty$ and the finial time
$\tau = \infty$, the system must be at one of the vacua labeled
by the integers \eqref{vacuaS2}.
Thus the action is evaluated as%
\footnote{More generic instanton solutions may be obtained
from the vacua with $\Phi = f = 1/g_{YM}(l_1,\cdots,l_N)$, 
where $l_i \in {\mathbb Z}$ is not restricted to the range $0 \leq l_i < k$.
Even for these generic vacua, we can construct instantons on 
$R_\tau \times S^2$ and therefore on $R_\tau \times S^3/{\mathbb Z}_k$ 
as well by utilizing the map of vacua.
If we want to use the range $0 \leq l_i < k-1$ for
${\cal A}_\chi = 1/g_{YM}(l_1,\cdots,l_N)$, then we 
just have to perform large gauge transformations.}
\begin{align}
 S =
       \frac{8 \pi^2}{k g_{\rm YM}^2}
       \left[ \sum_{I=0}^{k-1} n_I I^2 |_{\tau=\infty}
        -  \sum_{I=0}^{k-1} n_I I^2 |_{\tau=-\infty}  \right] ~.
        \label{instaction}
\end{align}
The possible interpolations of vacua were discussed in \cite{Lin}
by using the results of \cite{BHP}. 
In the dual gravity description, the amplitude $P \sim \exp ( - S)$ 
may be interpreted as the transition probability between geometries 
in the small AdS radius limit $l \to 0$.


\section{Conclusion and discussions}
\label{conclusion}

In this paper we have investigated the condensation of 
localized closed string tachyons in AdS orbifolds and 
its dual gauge theory description from the viewpoint of 
AdS/CFT correspondence.
The orbifolds of AdS space have fixed points at the center
and we can construct configurations with tachyonic modes
localized at the fixed points. The condensation of 
localized tachyon leads to the decay of AdS orbifolds
into more stable geometries. The dual theories are
given by orbifold gauge theories, and vacuum transitions of 
gauge theory correspond to geometry transitions of 
dual gravity theory.

As explicit examples, we have considered the orbifolds
of $AdS_{3}$ and  $AdS_5$ in type IIB superstring theory.  
First we have studied the localized tachyon 
condensation of $AdS_3/{\mathbb Z}_k$ with odd $k$.
The tachyon condensation leads to $AdS_3/{\mathbb Z}_{k'}$
with a smaller odd $k'$ and finally to pure $AdS_3$.
Assuming that the effect of localized tachyon condensation 
induces a dilaton pulse, we have constructed numerically a 
graviton-dilaton solution describing the decay.
The dual gauge theory description has been analyzed, and 
remarkably the Casimir energies are found to be the same as
the masses of dual geometries.
Then we move to the case of $AdS_5/{\mathbb Z}_k$, where the
AdS orbifold decays into Eguchi-Hanson soliton \cite{CM1,CM2}
after the localized tachyon condensation.
The gauge theory vacua dual to these geometries may have 
holonomies along the non-trivial cycle, and we have constructed 
instanton solutions interpolating different vacua as
non-perturbative transitions.

There are many interesting points to be investigated furthermore. 
As for the dynamics of the tachyon condensation in the gravity description, 
we could follow the time evolution before the dilaton pulse 
reaches the boundary. It is interesting 
to see how the solution behaves at the quite late time 
when the effects of boundary are significant. 
It is also true that the geometry changes can be induced
both by the condensation of tachyonic mode as perturbative 
effects of string theory and also by gravitational instantons
as non-perturbative effects. 
The localized tachyon condensation has been discussed in 
subsection \ref{gravity}, but the non-perturbative transition has
not been analyzed yet. This should be described by a gravitational
instanton which interpolate $AdS_3/{\mathbb Z}_k$ at $\tau = - \infty$
to $AdS_3/{\mathbb Z}_{k'}$ at $\tau = \infty$.
It is also important to analyze the $AdS_5$ case since the 
story is quite different from the $AdS_3$ case.

In the gauge theory description, the dynamics of vacuum
transition for (1+1) orbifold gauge theory is left to be
analyzed. However, we expect to obtain more insights
by studying deeply about the orbifold gauge theory on 
$R_t \times S^3/{\mathbb Z}_k$. 
We have examined non-perturbative effects in the gauge
theory description, but one may ask how to see the effect 
of localized tachyon condensation in this side.
It is actually a very difficult question as mentioned in \cite{APS}
because we are considering in the different regime of 't Hooft coupling.
The localized tachyon condensation has been investigated from
the viewpoint of dual gauge theory in \cite{AS,DKR1,DKR2}, 
but it is fair to say that no clear picture has been obtained yet.
The investigation in our configuration might give a clue since we
know the end point of tachyon condensation.

One of the main results of this paper is to extend the
analysis of $AdS_5$ case in \cite{Hikida} into the $AdS_3$ case.
In fact, the $AdS_3$ case could be more interesting since
we can solve string theory on $AdS_3$ with NSNS-flux
and go beyond the classical limit. 
For example, we can construct localized tachyons explicitly
as in \cite{MM1,MM2}, and it is also possible to analyze them
from the viewpoint of dual CFT.
Moreover, it is worthwhile trying to follow the 
RG flow of worldsheet theory in $AdS_3/{\mathbb Z}_k$, 
since the worldsheet RG flow leads important developments on 
the localized tachyon condensation \cite{APS,GHMS}. 
Tachyon condensations in string theory on $AdS_3$ have been also
discussed recently in \cite{LMT,BKR,RR} in different contexts.

\subsection*{Acknowledgement}

Y.H. would like to thank S.~Hirano, Y.~Okawa, I.~Papadimitriou, 
S.-J.~Rey and K.~Yoshida for useful discussions and KEK for its 
hospitality. The work of Y.H. was supported by JSPS Postdoctoral 
Fellowships for Research Abroad under the contract number H18-143.
N.I. would like to thank S. Hartnoll for fruitful discussion and 
H. Kudoh for many very helpful conversation about numerical analysis, and
especially S. Minwalla for wonderful collaboration on related works on 
AdS$_3$ tachyon while he was at TIFR. 
N.I. would also like to thank  
the string theory group at Caltech, Columbia University and 
University of Michigan, Ann Arbor for their very nice hospitality where 
part of this work was done. The research of N.I. was supported in part 
by the National Science Foundation under Grant No. PHY05-51164. 





\end{document}